\documentclass[twocolumn,showpacs, showkeys,preprintnumbers,amsmath,amssymb, prb]{revtex4}
\usepackage{graphicx}
\usepackage{dcolumn}
\usepackage{bm}
\usepackage[dvips]{color}
\begin{document}


\title{Magnetic properties of EuPtSi$_3$ single crystals}
\author{Neeraj Kumar}
\author{S. K. Dhar}
\affiliation{Department of Condensed Matter Physics and Material
Sciences, Tata Institute of Fundamental Research, Homi Bhabha Road,
Colaba, Mumbai 400 005, India.}
\author{P. Bonville}
\affiliation{2DSM/DRECAM/SPEC, CEA Saclay, 91191 Gif-Sur-Yvette, France}
\author{P. Manfrinetti}
\affiliation{Dipartimento di Chimica e Chimica Industriale, Universit\'{a} di Genova, Via Dodecaneso 31, 16146 Genova, Italy}
\author{A. Thamizhavel}
\email{thamizh@tifr.res.in}
\affiliation{Department of Condensed Matter Physics and Material
Sciences, Tata Institute of Fundamental Research, Homi Bhabha Road,
Colaba, Mumbai 400 005, India.}
\date{\today}

\begin{abstract}

Single crystals of EuPtSi$_3$, which crystallize in the BaNiSn$_3$-type  crystal structure, have been grown by high temperature solution growth method using molten Sn as the solvent.  EuPtSi$_3$ which lacks the inversion symmetry and has only one Eu site in the unit cell is found to be an antiferromagnet with two successive magnetic transitions at $T_{\rm N1}$ = 17~K and $T_{\rm N2}$ = 16~K, as inferred from magnetic susceptibility, heat capacity and $^{151}$Eu M\"{o}ssbauer measurements.  The isothermal magnetization data for $H~\parallel$~[001] reveal a metamagnetic transition at a critical field  $H_{\rm c}$ = 1~T.  The magnetization saturates to a moment value of  6.43~$\mu_{\rm B}$/Eu above 5.9~T ( 9.2~T) for $H~\parallel~$~[001] ( [100]),  indicating that these fields are spin-flip fields for the  divalent Eu moments along the two axes.  The origin of this anisotropic behaviour is discussed.   A magnetic (H, T) phase diagram  has been constructed from the temperature dependence of isothermal magnetization data.   The reduced jump in the heat capacity at $T_{\rm N1}$ indicates a transition to an incommensurate, amplitude modulated antiferromagnetic structure.  The shape of the hyperfine field split  M\"{o}ssbauer spectrum at $T_{\rm N1}$ provides additional support for the proposed nature of this magnetic transition.

\end{abstract}



\maketitle

\section{Introduction}

The discovery of superconductivity in the doped-$A$Fe$_2$As$_2$ ($A$ = Ba, Ca, Sr and Eu) spin density wave compounds has generated a great deal of activity  in search of new compounds, particularly with Ba, Ca, Sr and Eu  which exhibit interesting magnetic and superconducting properties~\cite{Neeraj, Goldman, Sefat, Ren}. Very recently, Bauer \textit{et al} have reported superconductivity at 2.25~K in BaPtSi$_3$ which crystallizes in the tetragonal BaNiSn$_3$-type non-centrosymmetric crystal structure with the space group $I4mm$~\cite{Bauer1}. Compounds possessing  a non-centrosymmetric crystal structure have been studied quite extensively after the discovery of superconductivity in CePt$_3$Si ($T_{\rm c}$ = 0.75~K)~\cite{Bauer2}. Several other Ce compounds with the general formula Ce$T$Si$_3$, where $T$ is a transition metal (Rh, Ir and Co) crystallizing in the BaNiSn$_3$-type crystal structure were later found to exhibit superconductivity under pressure~\cite{Kimura, Sugitani, Kawai1, Settai}.   The anisotropic magnetic properties of CePtSi$_3$ were studied by Kawai~\textit{et al} and they observed that the Ce moments order antiferromagnetically with two successive transitions at 4.8~K and 2.4~K respectively~\cite{kawai2}.  In this work, we report on the crystal growth, electrical and magnetic properties of EuPtSi$_3$ which also crystallizes in the BaNiSn$_3$ type crystal structure.  Interest in the Eu compounds stems from their unusual magnetic properties like valence fluctuation or magnetic ordering due to the strong Coloumb interaction between the $4f$ and conduction electrons.  Eu compounds are also appealing as the charged state and the magnetic interactions of the Eu ions can be probed by the technique of $^{151}$Eu M\"{o}ssbauer spectroscopy, supplementing the information derived from the usual bulk techniques like magnetization, resistivity and heat capacity.    We find that the magnetic properties of EuPtSi$_3$ show an unexpected anisotropy (Eu$^{2+}$ is an $S$-state ion, with $L$ = 0), whose origin is discussed.  The specific heat and M\"{o}ssbauer data suggest that two magnetic transitions take place at $T_{\rm N1}$ = 17~K and $T_{\rm N2}$ = 16~K.  The transition at 17~K is a paramagnetic incommensurate magnetic transition followed by a lock-in transition to a single moment, commensurate phase at 16~K; a phenomenon which is encountered in other Eu$^{2+}$ or Gd$^{3+}$ compounds.  

\section{Experimental}

The single crystals of EuPtSi$_3$ were grown by the high temperature solution growth using the fourth element, Sn as the solvent as reported by Kawai~\textit{et al}~\cite{kawai2} for the crystal growth of CePtSi$_3$. We took  the charge of high purity individual metals  Eu, Pt, Si and Sn in the ratio 1:1:3:19. The metals were placed in a recrystallized alumina crucible and sealed in a quartz ampoule with a partial pressure of argon gas.  The sealed crucible was then slowly heated to 1050~$^{\circ}$C  and kept at this temperature for one day to achieve proper homogenization.  Then the temperature of the furnace was cooled down to 500~$^{\circ}$C over a period of 3 weeks.  The grown crystals were extracted out of the Sn solvent  by means of centrifuging.  Several platelet like single crystals with typical dimension 4~$\times$3~$\times$~0.6~mm$^3$ were obtained.  The [001] axis is perpendicular to the plane of crystals.  Our attempts to make CaPtSi$_3$ by this method were unsuccessful and resulted in the formation of good single crystals of Ca$_2$Pt$_3$Si$_5$ which has recently been reported by Takeuchi \textit{et al}~\cite{Takeuchi}.  A polycrystalline sample of LaPtSi$_3$, as reference, non-magnetic analogue of the Eu compound, was prepared by the usual method of arc melting in an inert atmosphere of argon and annealed at 900~$^\circ$C for two weeks.

The dc magnetic susceptibility and the in-field magnetization measurements were performed in the temperature range 1.8\--300~K using a Quantum Design superconducting quantum interference device (SQUID) magnetometer and Oxford vibrating sample magnetometer (VSM). The temperature dependence of electrical resistivity in the range 1.8\--300~K was measured using a home made dc electrical resistivity set up. The heat capacity and magnetoresistance measurements were performed using a Quantum Design physical properties measurement system. $^{151}$Eu M\"{o}ssbauer spectra were recorded at various temperatures using a conventional acceleration spectrometer with a $^{151}$SmF$_3$ source.  Laue diffraction spots were recorded on a Huber Laue diffractometer, while powder diffraction pattern was recorded using  Phillips Pan-Analyitcal set-up.  Electron probe micro-analysis (EPMA) measurements were performed on a CAMECA SX100 electron microprobe.

\section{Results and Discussion}

In order to study the magnetic properties along the principal crystallographic directions, the single crystal of EuPtSi$_3$ was subjected to Laue diffraction; the good quality of the single crystals with four fold symmetry was confirmed by  Laue diffraction spots.  The flat plane of the crystal corresponded to the (001) plane.  The crystal was then cut along the principal crystallographic directions by means of spark erosion cutting machine. Since the single crystal was grown using Sn as flux, a few pieces of the single crystals were crushed into powder form and subjected to powder x-ray diffraction to check the phase purity of the sample.  From the Rietveld analysis, it is confirmed that the sample possesses tetragonal crystal structure with the space group $I4mm$.  The estimated lattice constants are $a$~=~4.2660~\AA ~and $c$ = 9.8768~\AA~with the unit cell volume 179.74~\AA$^3$.   Weak intensity peaks corresponding to free Sn were also detected in the powder spectra even after the surface of the crystal was thoroughly scrapped to remove any superficial Sn on the surface.  EuPtSi$_3$ has a layered structure and a thin layer of free Sn may get incorporated between the layers during the process of crystal growth.  The powder X-ray diffraction pattern of polycrystalline LaPtSi$_3$ was similar to that of EuPtSi$_3$ showing thereby that the two are iso-structural.  The lattice parameters obtained from the Rietveld analysis are $a$~=~4.3467~\AA~and $c$~=~9.6335~\AA~with the unit cell volume 182.01~\AA$^3$.  EPMA of the single crystals averaged over 10 different spots revealed the following composition~:~Eu$_{21.13}$Pt$_{20.50}$Si$_{58.29}$ : Sn$_{0.08}$ which is very close to 1:1:3 stoichiometry.  The trace amount of Sn present in the single crystals is unavoidable.

\begin{figure}[h]
\includegraphics[width=0.45\textwidth]{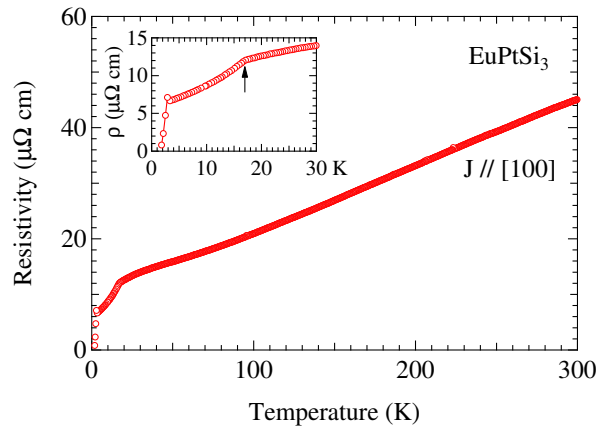}
\caption{\label{fig1}(Color online) Temperature dependence of electrical resistivity for $J~\parallel~$  [100] in the temperature range from 1.8 to 300~K.  The inset shows the low temperature part and the arrow indicates the magnetic ordering.}
\end{figure}

The temperature dependence of the electrical resistivity of EuPtSi$_3$  for  current parallel to [100] direction is shown in Fig.~\ref{fig1}.  The magnitude of electrical resistivity is  typical of intermetallic compounds and its decrease with decreasing temperature indicates a metallic state.   At 17~K the resistivity shows a change of slope (shown on an enlarged scale in the inset of Fig.~\ref{fig1}) due to reduction in  spin disorder scattering arising from magnetic ordering of the Eu moments .  The resistivity shows only one transition; however a second magnetic transition  is clearly seen in the magnetic susceptibility and the heat capacity data as described in the following.   Near $T$~=~3.7~K which is close to the superconducting transition temperature of Sn, the resistivity shows a drop which does not however attain a zero value.  This indicates the presence of some minor filamentary Sn incorporated in the crystal during  growth process.  

\begin{figure}[h]
\includegraphics[width=0.45\textwidth]{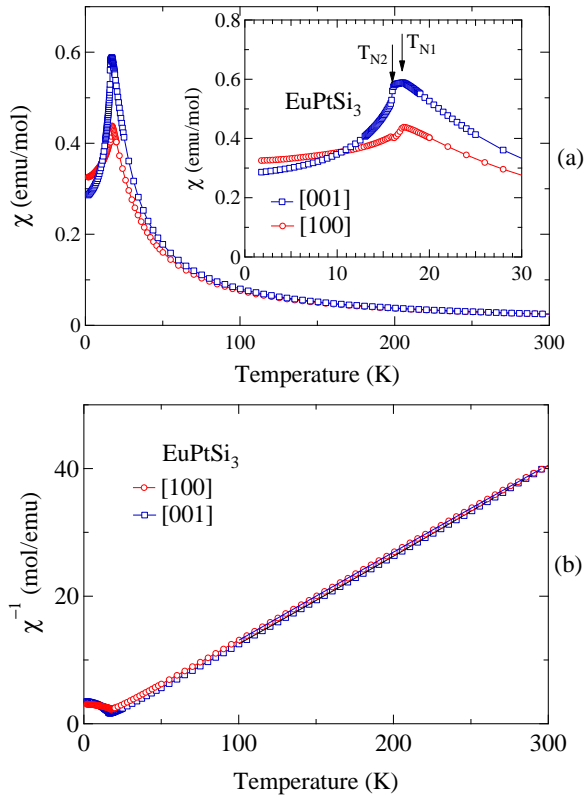}
\caption{\label{fig2}(Color online)(a) Temperature dependence of magnetic susceptibility along the two principal directions in the temperature range from 1.8 to 300~K measured in an applied field of 0.1~T.  The inset in (a) shows the low temperature part of the magnetic susceptibility clearly indicating the two antiferromagnetic ordering in EuPtSi$_3$.  (b) Inverse magnetic susceptibility of EuPtSi$_3$, the solid lines indicate the fit to Curie-Weiss law.}
\end{figure}

The temperature dependence of  magnetic susceptibility from 1.8 to 300~K is shown in Fig.~\ref{fig2} with the applied magnetic field $H$ parallel to [100] and [001] directions, respectively.   The low temperature magnetic susceptibility data clearly show two magnetic transitions at $T_{\rm N}$~=~17~K and 16~K.    The inverse magnetic susceptibility of EuPtSi$_3$ is shown in Fig.~\ref{fig2}(b) and the high temperature data were fitted to the Curie-Weiss law, $\chi = C/(T - \theta_{\rm p}$).  For $H~\parallel$~[100], the   effective moment  $\mu_{\rm eff}$ is 7.62~$\mu_{\rm B}$/Eu and the paramagnetic Curie temperature $\theta_{\rm p}$ = 4.7~K; for $H~\parallel$~[001], $\mu_{\rm eff}$ =  7.56~$\mu_{\rm B}$/Eu and $\theta_{\rm p}$ = 11.7~K. The effective moments are lower than the free ion value for Eu$^{2+}$ ($\mu_{\rm eff}$ =  7.94~$\mu_{\rm B}$), yielding a magnetic signal which is 8\% lower than expected.  Since we find the same signal deficit in the magnetization data, to be described below, we think it may be ascribed either to the presence of traces of an extra phase (Sn flux), resulting in a small over estimation of the actual amount of EuPtSi$_3$ present in the sample, or to a slight off-stoichiometry of Eu.  Evidence for the presence of Sn in the crystals, evoked above, favours the first possibility.  Since the magnetization data described below show that the magnetic structure is antiferromagnetic, the positive values of the paramagnetic Curie-Weiss temperature along both directions are \textit{a priori} surprising; they are tentatively attributed to exchange anisotropy,  as discussed below.  

\begin{figure}[h]
\includegraphics[width=0.45\textwidth]{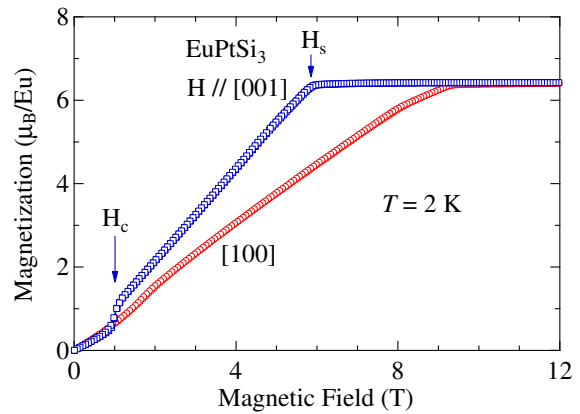}
\caption{\label{fig3}(Color online) Isothermal magnetization measurement of EuPtSi$_3$ along the two principal crystallographic directions.  The metamagnetic transitions are indicated by the arrows.}
\end{figure}

To further elucidate the nature of magnetic ordering, we have performed isothermal magnetization versus field scans at T = 2~K in a vibrating sample magnetometer up to a field of 12~T as shown in Fig.~\ref{fig3}.  At low field, the magnetizations along the two principal directions increase linearly with the field, with similar slopes.     For $H~\parallel$~ [001], there is an abrupt increase of the magnetization at 1~T followed by a faster linear increase up to 5.9~T; above 5.9~T, the magnetization is saturated with a value  6.43~$\mu_{\rm B}$/Eu.   For $H~\parallel$~[100],  the magnetization shows a linear increase with field and saturates to the same value at a higher field of 9.2~T. The apparent saturation moment  is lower than the Eu$^{2+}$ free ion value (7~$\mu_{\rm B}$/Eu) by 8\%, which, as discussed above, leads us to think that there is a small amount of Sn flux in the sample.  The overall behaviour of the magnetization versus field indicates an antiferromagnetic (AF) structure.  In a first analysis, the metamagnetic transition at $H_{\rm c}$~=~1~T for $H~\parallel~$ [001] and the smaller value for the saturation field along  [001] are characteristic of a two-sublattice AF structure, with moments in zero field along [001] and with a small crystalline anisotropy along this axis~\cite{Herpin}.  Then, the field $H_{\rm c}$ is the spin-flop field for which the moments along [001] reorient perpendicular to the field, and the critical fields of 5.9 and 9.2~T are the spin-flip fields ($H_{\rm s}$) for full alignment with the applied field in each direction.  The spin-flip field $H_{\rm s}$ along [100] is larger because [100] is a hard magnetic axis.  However, the fact that the magnetization along [001] is not zero below $H_{\rm c}$ and the presence of a slight metamagnetic transition along [100]  at 1.5~T show that the magnetic structure is more complex than described above, and in particular that the Eu$^{2+}$ moments (in zero field) are not exactly aligned along [001].  Anisotropy in the spin flip fields has been observed in some Gd compounds~\cite{Garnier}, in EuAs$_3$~\cite{Bauhofer} and in EuPdSb~\cite{Bonville}.  It arises in principle from the crystal field~\cite{Herpin}, although typically Eu$^{2+}$ and Gd$^{3+}$ ions, which have $L$~=~0, show very small magneto-crystalline anisotropy.  Another source of anisotropy could be two-ion exchange, although the simple model of an anisotropic molecular field tensor developed in Ref.~\cite{Bonville} for a two-sublattice AF structure predicts that the spin-flip fields should be the same for both directions.  So we tentatively attribute the observed behaviour mainly to an anomalously large crystalline anisotropy of Eu$^{2+}$ in EuPtSi$_3$.    
More quantitatively, the two-sublattice model of Ref.\cite{Herpin} involves the exchange field $H_{\rm e}$ and the anistropy field $H_{\rm A} = K/m_0$, where $K$ is the crystalline anisotropy energy and $m_0$ the saturated Eu$^{2+}$ moment.  Along the easy axis [001], the spin-flop field is $H_{\rm c}$ = $2\sqrt{H_{\rm A}(H_{\rm e} - H_{\rm A}})$  and the spin-flip field $2(H_{\rm e} - H_{\rm A})$, and along [100] the spin flip field is $2(H_{\rm e} + H_{\rm A})$.  Applying the model to EuPtSi$_3$, the values of the critical fields along [001] can be reproduced with $H_{\rm e}$ = 3.0~T and $H_{\rm A}$ = 0.085~T, which are reasonable values, but then the critical field along [100] is 6.25~T, too small with respect to experiment (9.2~T).  So this simple model also fails to reproduce in detail the magnetization behaviour, which is probably due to a magnetic structure more complex than two AF sublattices with moments aligned along [001].  Nevertheless, a hint to the presence of exchange anisotropy is given by the different values of the paramagnetic Curie temperature along the two axes.  Indeed, in terms of the exchange integrals with nearest neighbours $\mathcal{J_{\rm 1i}}$ and with second neighbours $\mathcal{J_{\rm 2i}}$ along direction $i$, one has, on the one hand~\cite{Blanco}:
\begin{equation}
k_{\rm B} \theta_{\rm pi} = \frac{2S(S+1)}{3} \left(\mathcal{J_{\rm 1i}} + \mathcal{J_{\rm 2i}}\right).
\end{equation}
The anisotropy of $\theta_{\rm p}$ then can stem from an anisotropy of the exchange constants.  On the other hand, one has:
\begin{equation}
k_{\rm B} T_{\rm N} = \frac{2S(S+1)}{3} \left|\mathcal{-J_{\rm 1k}} + \mathcal{J_{\rm 2k}}\right|,
\end{equation}
where $k$ labels the direction of the propagation vector of the magnetic structure.  In case $\mathcal{J_{\rm 1k}}$ is negative and $\mathcal{J_{\rm 2k}}$ is positive with $\mathcal{J_{\rm 2k}} > |\mathcal{J_{\rm 1k}}|$, then the magnetic structure is of AF type.  If these conditions are also fulfilled for the exchange integrals along directions [001] and [100], then the paramagnetic Curie temperatures are positive, and $T_{\rm N} > \theta_{\rm pi}$, as observed EuPtSi$_3$.  

\begin{figure}[h]
\includegraphics[width=0.45\textwidth]{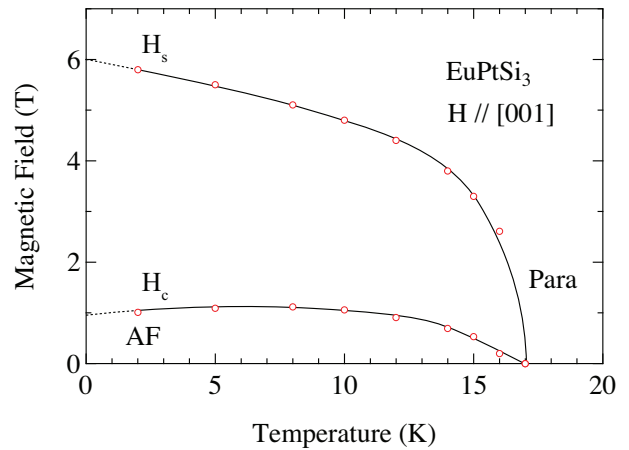}
\caption{\label{fig4}(Color online) Magnetic phase diagram of EuPtSi$_3$ determined from the isothermal magnetization measured at various temperature. }
\end{figure}

We have also recorded the magnetization data at various temperatures for $H~\parallel~$ [001].  As the temperature is increased, the spin-flop transition shifts slightly to higher fields and then it decreases as the temperature is increased.  The spin-flip field monotonically decreases as temperature is increased.  From the $dM/dH$ plots (not shown here) we have constructed the magnetic phase diagram as shown in Fig.~\ref{fig4}.

\begin{figure}[t]
\includegraphics[width=0.45\textwidth]{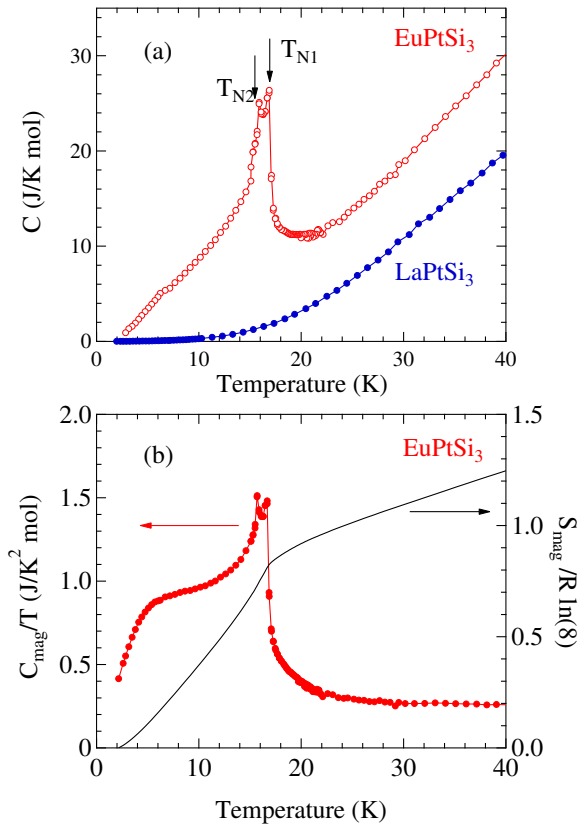}
\caption{\label{fig5}(Color online)(a)  Temperature dependence of the heat capacity of EuPtSi$_3$.  The two magnetic orderings are indicated by the arrow. (b)  $C/T$ versus $T$ of EuPtSi$_3$, the calculated entropy is shown as solid line.}
\end{figure}

Figure~\ref{fig5} shows the temperature dependence of heat capacity of EuPtSi$_3$ measured between 2 and 40~K, together with the heat capacity of non-magnetic, reference polycrystalline LaPtSi$_3$.  The heat capacity clearly shows two peaks at 17~K and 16~K confirming the bulk magnetic ordering in this compound. Incidentally, isostructural CePtSi$_3$ also exhibits two magnetic transitions at 4.8~K and 2.4~K respectively~\cite{kawai2}.  The magnetic part of   heat capacity $C_{\rm mag}$  was deduced by the usual method of subtracting the heat capacity of LaPtSi$_3$ from that of EuPtSi$_3$ after taking into account the renormalization due to different atomic masses of La and Eu.  The $C_{\rm mag}/T$ versus temperature plot and the calculated entropy is shown in Fig.~\ref{fig5}(b). The entropy is about $0.8~R$~ln8 at $T_{\rm N1}$  but above $T_{\rm N1}$  it exceeds the theoretical value of $R$~ln8 for $S = 7/2$. While short range order above $T_{\rm N1}$, which is typically present in the paramagnetic state, would explain the observed values of entropy up to $T_{\rm N1}$, the increase of entropy beyond $R$~ln8 at higher temperatures suggests that LaPtSi$_3$  may not be a good reference for the lattice heat capacity of EuPtSi$_3$. We believe such an indication is reflected by the heat capacity plots of the two compounds which show an increasing divergence from each other at high temperatures. The different valence states of Eu$^{2+}$ and La$^{3+}$ may result in different inter-atomic potentials modifying the phonon spectra of these two compounds.   The heat capacity jump $\Delta C_{\rm mag}$ at $T_{\rm N}$ is estimated to be 14.6~J/K mol. Based on the mean field approximation, Blanco~\textit{et al}~\cite{Blanco} have calculated the heat capacity jumps at the magnetic transition for two types of magnetic structures, namely, the equal moment (EM)  where the magnetic moments are the same at all sites and  the amplitude modulated (AM), where the magnetic moment amplitude varies periodically from one site to another.   For the case of EM structure, the jump in the heat capacity at the ordering temperature is given by,
\begin{equation}
\Delta C_{\rm EM} = 5 \frac{J(J+1)}{(2J^2 + 2J + 1)} R
\end{equation}
and for the amplitude modulated system,
\begin{equation}
\Delta C_{\rm AM} = \frac{10}{3} \frac{J(J+1)}{(2J^2 + 2J + 1)} R
\end{equation}

where $J$ is the total angular momentum.  Since the effects of crystal electric field for $L = 0$, divalent state of Eu are negligible, the total angular momentum $J$~ (=7/2) can be used in the above equations.  $\Delta C_{\rm EM}$ amounts to 20.14~J/K$\cdot$mol for the EM case, while for the AM case, $\Delta C_{\rm AM}$ equals 13.4~J/K$\cdot$mol. Thus, theoretically, the jump in the heat capacity at the magnetic transition for AM case is reduced to 2/3 of the value for EM case~\cite{Blanco}.  Our estimate of $\Delta C_{\rm mag}$ (14.6~J/K$\cdot$mol) at $T_{\rm N1}$  thus  strongly suggests that EuPtSi$_3$ possesses an amplitude modulated structure.  Furthermore, Blanco~\textit{et al} found that for the amplitude modulated system a hump in the heat capacity occurs below the ordering temperature, to compensate the loss of entropy just below $T_{\rm N}$, which incidentally is seen in EuPtSi$_3$ as shown in Fig.~\ref{fig5}(b) in the $C/T$ versus temperature plot.

\begin{figure}[h]
\includegraphics[width=0.35\textwidth]{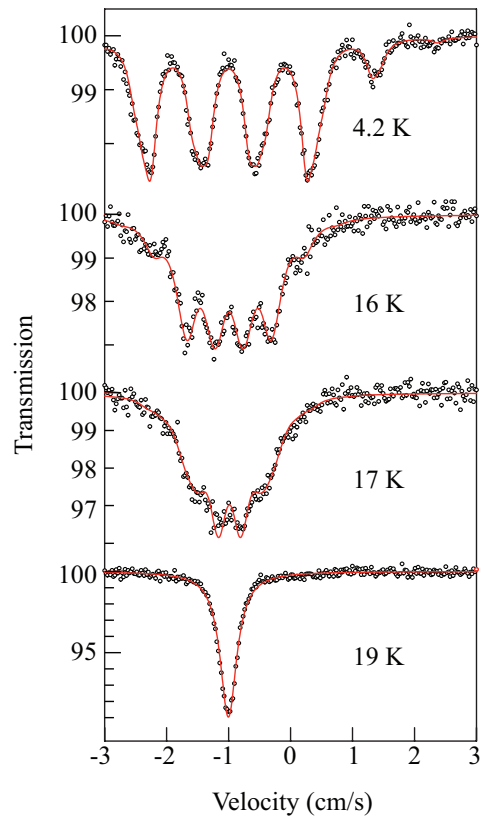}
\caption{\label{fig6}(Color online) $^{151}$Eu Mossbauer absorption spectra at selected temperatures in EuPtSi$_3$.  The lines are fits to a single magnetic hyperfine interaction at 4.2 and 16~K, to an inconmmensurate magnetic modulation at 17~K and to a sinlge Eu$^{2+}$ line at 19~K.}
\end{figure}

$^{151}$Eu M\"{o}ssbauer spectroscopy is an excellent tool to probe the valence and the magnetic state of Eu ions in Eu compounds. While information about the valence state of Eu ions is obtained via the isomer shift, the magnitude of the hyperfine magnetic field can be inferred from the splitting of the spectrum due to hyperfine interaction. $^{151}$Eu M\"{o}ssbauer absorption spectra were recorded in EuPtSi$_3$ in the temperature range 4.2 to 19~K and at 300~K.  Spectra at selected temperatures are shown in Fig.~\ref{fig6}.  All the spectra point to the presence of Eu$^{2+}$ with no traces of Eu$^{3+}$ impurity content.  The isomer shift of all the spectra was found to be -10.0(4) mm/s arising from the divalent state of Eu ions.   A single line is observed from room temperature down to 19~K.   At 300~K the line width is 2.65~mm/s which is a standard value for a  Eu$^{2+}$ compound and it increases to 3.2~mm/s at 19~K, due to the dynamic short range order close to the first magnetic transition at $T_{\rm N}$ = 17~K.  At 4.2~K, a magnetic hyperfine field of 33.0(5)~T can be derived from the spectrum, which is a standard value for the Fermi contact interaction in Eu$^{2+}$ materials.  On increasing the temperature, the hyperfine field decreases steadily, reaching 16.8(5)~T at 16~K and the lines are narrow, indicative of a commensurate magnetic order with a single moment value (EM structure).  At 17~K, the spectral shape changes suddenly, as can be seen in the Fig.~\ref{fig6}: a broad distribution of hyperfine fields is present at the nucleus sites.  The spectrum can be fitted to a Gaussian-shaped distribution, but a physically more appealing interpretation is the following: below 17~K, where the specific heat data shows its first anomaly, an incommensurate AM magnetic structure develops, which yields a distribution of Eu$^{2+}$ moments, and hence of hyperfine fields.   The second peak at 16~K in the specific heat would correspond to a lock-in transition to a commensurate wave-vector, which is not uncommon in intermetallic Eu or Gd compounds~\cite{Blanco}.    The fit at 17~K shown in the figure corresponds to  an incommensurate modulation, but it cannot be distinguished from that to a Gaussian shaped distribution.  However, it is difficult to understand why a broad Gaussian distribution of hyperfine magnetic fields would appear at 17~K, if it were not because of a phase transition yielding a specific moment distribution.  Indeed, it has been reported that such a ``cascade'' of transitions (paramagnetic - incommensurate AM - commensurate EM) occurs for instance in layered semi-metallic EuAs$_3$~\cite{Chattopadhyay} below $T_{\rm N1}$ = 11.3~K and $T_{\rm N2}$ = 10.26~K, as well as in EuPdSb~\cite{Bonville}, below $T_{\rm N1}$ = 18~K and $T_{\rm N2}$ = 12~K.  We therefore believe the  M\"{o}ssbauer and the heat capacity data corroborate each other quite well in demonstrating the presence of a ``cascade'' of transitions in EuPtSi$_{3}$.

\section{Conclusion}

Single crystals of EuPtSi$_3$ were grown by flux method and the magnetic studies revealed antiferromagnetic transitions at $T_{\rm N1}$ = 17 and $T_{\rm N2}$ = 16~K.  From the low temperature isothermal magnetization measurements, we found a spin-flop-like transition at $H_{\rm c}$ = 1~T for  $H~\parallel$~[001]  and spin-flip transitions at 5.9~T (9.2~T) for $H~\parallel$~[001] ( [100]).  This shows the presence of anisotropy, probably of both crystalline and exchange origin.   A magnetic phase diagram has been constructed based on the isothermal magnetization data.  The heat capacity measurement also confirm the occurrence of two magnetic transitions. Further,  the specific heat and M\"{o}ssbauer data  suggest that EuPtSi$_3$ undergoes a cascade of close transitions (paramagnetic - incommensurate - commensurate) as temperature  decreases.

\end{document}